\documentclass[a4paper,aps,pra,twocolumn,showpacs,preprintnumbers,amsmath,amssymb]{revtex4}

\usepackage{bbm}
\usepackage{amsmath}
\usepackage{verbatim}
\usepackage{graphicx}
\usepackage{epsfig}
\usepackage{times}

\newcommand{\beq}{\begin{equation}}
\newcommand{\eeq}{\end{equation}}
\newcommand{\beqa}{\begin{eqnarray}}
\newcommand{\eeqa}{\end{eqnarray}}

\newcommand{\ket} [1] {\vert #1 \rangle}
\newcommand{\bra} [1] {\langle #1 \vert}

\newcommand{\mean}[1]{\langle #1 \rangle}

\begin{document}
\title{On amplification of light in the continuous EPR state}

\author{V.N. Gorbachev,\footnote{E-mail: vn@vg3025.spb.edu}
A.I. Trubilko.}

\affiliation{Laboratory of Quantum Information and Computing,
University of AeroSpace Instrumentation, 67, Bolshaya Morskaya,
St.-Petersburg, 190000, Russia}

\begin{abstract}
Two schemes of amplification of two-mode squeezed light in the
continuous variable EPR-state are considered. They are based on
the integrals of motion, which allow conserving quantum
correlations whereas the power of each mode may increase. One of
these schemes involves a three-photon parametric process in a
nonlinear transparent medium and second is a Raman type
interaction of light with atomic ensemble.
\end{abstract}


\maketitle

%%%%%%%%%%%%%%%%%%%%%%%%%%%%%%%%%%%%
\section{Introduction}
Optics implementation of entangled states, particulary of the
continuous variable EPR type, is usually based on the light in
non-classical state known as squeezed light. Because of its
quantum correlations squeezed light is fragile and some special
transformations can only conserve its features. It is known, that
amplification of non-classical light is a hard problem. It has
been shown by many authors, see, for example Caves \cite{Caves},
Sokolov \cite{IVS}, that the usual linear optics amplifier is not
suitable because of its noise due from spontaneous emission. This
argument can be extended into nonlinear amplifier \cite{OurEx} and
propagation through active media considered by Paris \cite{Paris}
with respect to quantum information tasks. In the same time the
non-classical states of light can be converted from one frequency
to another perfectly in a crystal with quadratic
nonlinearity \cite{VA}.\\
The aim of this paper is to consider amplification of two-mode
squeezed light of the EPR form, that can be generated in Optical
Parametric oscillator (OPO). Light of such type has been used for
teleportation of coherent state \cite{CWtel}, indeed its power was
week. The main idea is based on existence of  integrals of
motions, when light interacts with medium.  The reason is that
some entangled states can be eigenfunctions of the operators to be
integrals, it results in conservation of quantum correlations,
particulary entanglement, wile power of light may increase. Some
examples of interactions, for which the non-classical properties
of light unchange, have been discussed in ref.\cite{VA3}. An
thermostat, that creates and keeps atomic entanglement has been
introduced by Basharov \cite{Askhat}. However, when light
propagates trough  a phase-sensitive environment, its quantum
correlations degrades, as it has been found by Lee and co-workers
\cite{Lee}.

In this paper we introduce two schemes for amplification. First
involves a three-photon parametric process in transparent medium
with quadratic nonlinearity, second is a Raman-type interaction of
squeezed light with atomic ensemble. The peculiar property of
these schemes is that they can't generate entangled states but
they conserve degree of correlations between modes or
entanglement, wile the power of each modes may increase. It
directly results  from integrals of motion, which allow achieving
two tasks as conserving of properties we wish and appearance of a
process we wish. We consider the time evolution here, but the
presented method, based on integrals of motion, can be modified to
propagation problems. It can be done with the use of the quantum
transfer formalism, presented in ref. \cite{TransF}, where it has
been discussed some examples, that show how integrals of motion
work in parametric and multiphoton phenomena.

The paper is organized as follows. First some features of the
continuous variables EPR states are discussed, then amplification
of entangled light in nonlinear transparent medium and in
resonance medium is considered.

\section{Squeezed light and continuous EPR state}

There are two ways how to introduce the continuous analog of the
EPR pair. First of them includes formal definition in the
eigenfunction of two operators to be a total momentum $P$ and a
relative position $Q$ of a bipartite system. Second is based on a
physical model of  OPO, implemented in experiment. Its output is
entangled state of the EPR type, that can be described by
operators $P$ and $Q$ in the Heisenberg picture. In optics
implementation both ways result in the same state with respect to
measured values, when one finds a two-mode squeezed light.

Operators of total momentum and relative position of a bipartite
system, or two particles for simplicity, can be introduced in the
following way
\begin{eqnarray}
\label{PQ}
 \nonumber
Q=x_{1}-\epsilon x_{2},&&\\
P=p_{1}+(1/\epsilon )p_{2},&&
\end{eqnarray}
where $x_{m},p_{m}$, $m=1,2$ are the canonical position and
momentum operators of single particle, $\epsilon$ is a real
c-number. Operators $P$ and $Q$ have a set of common
eigenfunctions known as the continuous Bell-like states introduced
by Braunstein \cite{BrBell}
\begin{eqnarray}\
\label{BS} \nonumber
Q\ket{\Psi_{PQ}}=\mathcal{Q}\ket{\Psi_{PQ}},&&\\
P\ket{\Psi_{PQ}}=\mathcal{P}\ket{\Psi_{PQ}}.&&
\end{eqnarray}
When eigenvalues are equal to zero, there is a continuous analog
of maximally entangled EPR state
\begin{equation}\label{11}
\ket{\Psi_{00}}=\ket{EPR}.
\end{equation}

A simple model of OPO can be described by the effective
hamiltonian of interaction
$H=ik\hbar(a^{\dagger}_{1}a^{\dagger}_{2}-h.c.)$, where $k$ is a
coupling constant, $a_{m}$ is a photonic  operator of mode
$m=1,2$. Solution of the problem has the form
$Q=Q_{0}\exp(-\epsilon r)$, $P=P_{0}\exp(-\epsilon r)$, where
$\epsilon=\pm 1$, $Q_{0}, P_{0}$ are input operators, $r$ is a
squeezing parameter. When $r\to \infty$, one finds a continuous
variables EPR state, that can be denoted as \beqa \label{EPR}
\nonumber
Q\to 0,&&\\
P\to 0.&& \eeqa This is an ideal EPR pair, but it is not a good
physical state.

In quantum optics light can be described by the usual quadrature
operators
\begin{eqnarray}\label{12}
%\nonumber
 X(\theta)=
a^{\dagger}\exp(i\theta)+h.c.%&&\\
=2(x\cos\theta+ p\sin\theta),&&
\end{eqnarray}
where $a=x+ip$ is photonic operator, $[a;a^{\dagger}]=1$, $[x;
p]=i/2$. Any features of light can be obtained from a set of its
correlation functions, particulary from variances of the
quadrature operators $\mean{(\Delta
X)^{2}}=\mean{X^{2}}-\mean{X}^{2}$, which measurement are well
known. For example, in the case of coherent state $\mean{(\Delta
X)^{2}}=1$. If
 \beq
\label{120} \mean{(\Delta X(\theta))^{2}}<1, \eeq the light is
called squeezed or in more accuracy the state is squeezed over
position or amplitude, when $\theta=0$ and it is squeezed over
momentum or phase when $\theta=\pi/2$. If $\mean{(\Delta
X(\theta))^{2}}=0$ one finds a limit squeezing. Indeed, we use
terms amplitude and phase in the sense of consideration in phase
space, but not with respect to operators of amplitude and phase.

The introduced  quadrature operators are measured in a detection
scheme including  two detectors and a local field oscillator, that
is mixed with the signal. Output is the difference of the detector
photocurrents $i$. Assume, the quantum efficiency of both
detectors is equal to 1, then there is a simple formula for
spectrum of photocurrent or spectrum of the light noise. It is
described by variance of the quadrature operators and has the form
\begin{eqnarray}
\nonumber
i^{2}(\omega)= \nonumber
\int_{-\infty}^{\infty}d\tau\mean{i(t)i(t+\tau)}\exp(i\tau\omega)&&\\
=\int_{-\infty}^{\infty}d\tau\mean{X(t)X(t+\tau)}\exp(i\tau\omega).&&
\end{eqnarray}
Indeed, the large number of the quantum optics models results in
the spectrum of noise in the low region of frequency, that reads
\begin{equation}\label{13}
i^{2}(\omega\approx 0)=1+\mean{(\Delta
X(\theta))^{2}}_{\mathcal{N}},
\end{equation}
where unit is a level of shot noise known also as standard quantum
limit, subscript $\mathcal{N}$ denotes the normal ordering of
operators. It follows from (\ref{120}) and (\ref{13}), that
squeezed light has its noise bellow the shot level and it may be
suppressed even up to zero.

Operators $P$ and $Q$, given by (\ref{PQ}), can be rewritten in
terms of the quadrature operators in the following way
\begin{eqnarray}
\label{37T}
 \nonumber
  Q=(1/2)[X_{1}(0)-\epsilon X_{2}(0)],&&\\
P=(1/2)[X_{1}(\pi/2)-(1/\epsilon) X_{2}(\pi/2)],
\end{eqnarray}
where $X_{m}$, describes mode $m=1,2$. Both observables $P$ and
$Q$ are measured in a detection scheme, involved a beamsplitter,
that mixes two modes, and the next measurement of momentum and
position of light from two outputs of the beamsplitter
\cite{CWtel}. It follows from (\ref{BS}) and (\ref{EPR}), that
variances of $P$ and $Q$ are equal to zero, then equations
(\ref{120}) and (\ref{13}) tell, that the continuous Bell-state
$\ket{\Psi_{PQ}}$, together with output state of OPO, are squeezed
over total momentum and relative position and their noise is
suppressed below standard quantum limit. One finds here, that
these squeezed states are also entangled, but a simple observation
shows, that squeezing or better non-classicality of the light
state is only a necessary condition. In the same time, when light
is entangled, then the measure of entanglement can be chosen as a
level of the shot noise suppression.

\section{Amplification in parametric process}

Let an interaction of two modes be presented by hamiltonian of the
form \beqa \label{Ham} V=\hbar kPQ,
 \eeqa
where $k$ is a real coupling constant. Then operators of total
momentum $P$ and relative position $Q$ are integrals of motion,
that results in conservation of quantum correlations. Let the
initial state of light be an EPR pair generated by OPO, that has
the form $Z(t=0)=Z_{0}\exp(-\epsilon r)$, $Z=P,Q$. Then one finds,
that $Z(t)=Z(0)$. Also any eigenstates of integrals $Z=P,Q$ are
unchanged and entangled states $\ket{\Psi_{PQ}}$ are conserved,
because of their evolution is reduced to multiplication to a phase
factor: $\exp(-i\hbar^{-1}Vt)\ket{\Psi_{PQ}}=
\exp(-ik\mathcal{PQ}t)\ket{\Psi_{PQ}}$.

In the same time each of modes can be changed wile the
interaction.  The problem, given by (\ref{Ham}), has exact
solutions for the photon number operators of modes
$n_{m}=a^{\dagger}_{m}a_{m}$, $m=1,2$ \beqa \label{22}
 \nonumber
n_{1}(t)=n_{10}+\mu^{2}(Q^{2}+P^{2})+ 2\mu(Qx_{10}+Pp_{10}),
&&\\
\nonumber
n_{2}(t)=n_{20}+\mu^{2}(Q^{2}/\epsilon^{2}+P^{2}\epsilon^{2})&&\\
+ 2\mu(Qx_{20}/\epsilon+ Pp_{20}\epsilon),&& \eeqa where $Y_{m0}$,
$m=1,2$ are operators at $t=0$ and $\mu=kt/2$. In accordance with
(\ref{22}) the maximally entangled states $\ket{EPR}$, given by
(\ref{EPR}), is unchanged with respect to number of photons or its
power, because of $Q,P\approx 0$. For this case of ideal EPR pair
the medium plays a role of a quantum repeater. When the EPR pair
is not ideal its power may increase with a gain to be proportional
to $\mu^{2}$.

The natural question is whether the hamiltonian  (\ref{Ham}) can
describe any real process. To answer the question let rewrite the
hamiltonian in terms of photonic operators \beqa \label{ham}
\nonumber V=i\frac{k}{4}[a_{1}^{\dagger
2}-a_{1}^{2}-a_{2}^{\dagger 2}
+a_{2}^{2}&&\\
\nonumber
+(1/\epsilon)(a_{1}^{\dagger}a_{2}^{\dagger}-a_{1}a_{2})(1-\epsilon^{2})&&\\
+(1/\epsilon)(a_{1}a_{2}^{\dagger}-a_{1}^{\dagger}a_{2})(1+\epsilon^{2})].&&
\eeqa Assume $\epsilon=\pm 1$, then one finds three-photon
parametric phenomena in transparent medium with quadratic
nonlinearity. In fact there are three processes of the frequency
conversions, presented in (\ref{ham}). Two of them are down
conversions included the strong classical pumping
$\Omega_{m}=\omega_{m}+\omega_{m}$ and the reminder is up
conversion of the type $\Omega+\omega_{1}=\omega_{2}$, where
$\Omega_{m}, \Omega$ are frequencies of the pumping waves, and
$\omega_{m}$ is frequency of the mode $m=1,2$, associated with
operator $a_{m}$. All these processes may occur effectively in the
nonlinear medium discussed in ref. \cite{ASC},\cite{ASCh}.

\section{Amplification in resonance medium}

To consider a resonance interaction of light with an ensemble of
$N$ two-level identical atoms let introduce an effective
hamiltonian
\beqa \label{10} \nonumber H=i\hbar\vartheta&&\\
\vartheta=S_{10}B-S_{01}B^{\dagger},&&
 \eeqa
where atomic operators read $ S_{xy}=\sum_{a}s_{xy}(a)$,
$s_{xy}(a)=\ket{x}_{a}\bra{y}$, $x,y=0,1$, $\ket{0}_{a}$ and
$\ket{1}_{a}$ are ground and excited levels of an atom $a$. In the
hamiltonian e.m. field is presented by its operators $B,
B^{\dagger}$.

Assume, three modes $M=1,2,3$ at frequencies $\omega_{M}$,
 described by operators $a_{M}$, interact with
atoms by Raman type so that $\omega_{1}=\omega_{0}$ and
$\omega_{3}-\omega_{2}=\omega_{0}$, where $\omega_{0}$ is a
frequency of the atomic transition $0\to 1$. The operators $B$
takes the form $B=ga_{1}-fa_{3}a^{\dagger}_{2}$, where $g, f$ are
coupling constants. Let the mode at frequency $\omega_{3}$ be a
classical wave, then $B=g(a_{1}+\nu a^{\dagger}_{2})$, where
$\nu=fa_{3}/g$. Assuming $\nu=\epsilon=\pm 1$, the field operators
read
 \beqa \label{B}
 \nonumber
B=g(a_{1}-\epsilon a^{\dagger}_{2})=Q+iP,&&\\
B^{\dagger}=g(a_{1}^{\dagger}-\epsilon a_{2})=Q-iP.&& \eeqa
Equation (\ref{B}) tells, that operators $P$ and $Q$, where
$\epsilon^{2}=1$, are integrals of motion. Thanks to these
integrals quantum correlations of the entangled states of the EPR
type will be unchanged under evolution, given by hamiltonian
(\ref{10}), (\ref{B}).

To discuss amplification of modes $\omega_{1}, \omega_{2}$
introduce the master equation for density matrix of field $\rho$.
In the first approximation over hamiltonian of interaction it has
the form \beqa \label{Ku} \nonumber
\dot{\rho}=-\frac{N_{0}}{\gamma_{\bot}}(B^{\dagger}B\rho+
B\rho B^{\dagger}+h.c.)&&\\
-\frac{N_{1}}{\gamma_{\bot}}(BB^{\dagger}\rho+B^{\dagger}\rho
B+h.c.),&& \eeqa where $N_{0}$ and $N_{1}$ are occupations of the
ground and excited atomic levels, $N_{0}+N_{1}=N$ and  the
transversal decay rate $\gamma_{\bot}$ is introduced. It follows
from (\ref{Ku}), that equations for the photon numbers of single
modes have the form
 \beqa
 \nonumber
\mean{\dot{n}_{1}}=\frac{g^{2}}{\gamma_{\bot}}(N_{1}-N_{0})
(\mean{n_{1}}-\mean{n_{2}}+\mean{BB^{\dagger}})&&\\
\nonumber +\frac{g^{2}}{\gamma_{\bot}}(N_{0}+N_{1}),&&\\
%\nonumber
\mean{\dot{n}_{1}}-\mean{\dot{n}_{2}}=\frac{2g^{2}}{\gamma_{\bot}}
(N_{1}-N_{0})\mean{BB^{\dagger}}.&&%\\
%\nonumber
 %\mean{\dot{n}_{1}}+\mean{\dot{n}_{2}}=
%\frac{2g^{2}}{\gamma_{\bot}}[(N_{1}-N_{0})(\mean{n_{1}}&&\\
%-\mean{n_{2}})+N_{0}+N_{1}].&&
 \eeqa
There are two kind of amplification here.  If $N_{1}=N_{0}$, then
the power of each of modes increases with the gain to be
proportional to total number of atom $N$, but the relative power
of modes is unchanged. Let $\mean{BB^{\dagger}}=0$, it means,
that, for example, the initial state is EPR states for which $Q\to
0$, $P\to 0$. Then
$\mean{n_{m}}=n_{m0}+(g^{2}t/\gamma_{\bot})((N_{1}-N_{0})
(n_{10}-n_{20})+ N_{1}+N_{0})$. It results in amplification of
both modes wile its quantum correlations are conserved, if these
modes have the same number of photons initially.

%\section{Conclisions}

\begin{acknowledgments}
We acknowledge discussions with A. Basharov and S. Kulik. This
work was supported in part by the Delzell Foundation Inc. and
INTAS grant no. 00-479.
\end{acknowledgments}

%\bibliography{quantinf}

\end{document}